\documentclass[11pt]{article}
\usepackage[normalem]{ulem}
\usepackage{subfigure}
\usepackage{graphicx}
\usepackage{hyperref}
\usepackage{epsfig,subfigure}
\usepackage{color}
\usepackage{enumerate}

\usepackage{xcolor,cancel}

\usepackage[utf8]{inputenc}
\usepackage{epsfig,subfigure}
\usepackage[cmex10,fleqn]{amsmath}
\usepackage{graphicx}
\usepackage{color}
\usepackage{amssymb}
\usepackage{pifont}
\usepackage{hyperref}
\usepackage{verbatim}
\usepackage{stfloats}
\usepackage[bookmarks=false]{}

\def\QED{\mbox{\rule[0pt]{1.5ex}{1.5ex}}}
\def\proof{\noindent\hspace{2em}{\it Proof: }}

\newtheorem{theorem}{Theorem}
\newtheorem{corollary}{Corollary}
\newtheorem{definition}{Definition}

\begin{document}
\date{}
\title{Network Coherence Time Matters --\\ Aligned Image Sets and the Degrees of Freedom of  Interference Networks with Finite Precision CSIT and Perfect CSIR}
\author{ \normalsize Arash Gholami Davoodi and Syed A. Jafar \\
{\small Center for Pervasive Communications and Computing (CPCC)}\\
{\small University of California Irvine, Irvine, CA 92697}\\
{\small \it Email: \{gholamid, syed\}@uci.edu}
}
\maketitle


\allowdisplaybreaks
\begin{abstract}
This work obtains the first bound that is provably sensitive to \emph{network} coherence time, i.e., coherence time in an interference network where all channels experience the same coherence patterns. This is accomplished by a novel adaptation of the aligned image sets bound, and settles various open problems noted previously by  Naderi and Avestimehr  and by Gou et al. For example, a necessary and sufficient condition is obtained for the optimality of $1/2$ DoF per user in a partially connected interference network where the channel state information at the receivers (CSIR) is perfect, the channel state information at the transmitters (CSIT) is instantaneous but limited to finite precision, and the network coherence time is $T_c=1$. The surprising insight that emerges is that even with perfect CSIR and instantaneous finite precision CSIT, network coherence time matters, i.e., it has a DoF impact. 
\end{abstract}

\section{Introduction}
The impact of coherence time in a wireless network is a topic that has been studied extensively \cite{Hassibi_Hochwald,  Lapidoth,  Lapidoth_Shamai_Wigger_BC, Jafar_mobile, Jafar_corr, CBIA, Wang_Gou_Jafar, Jafar_TIM, Naderi_Avestimehr}.  Nevertheless some of the most fundamental questions about coherence remain unanswered.  For example, it is well known that longer coherence time is beneficial to amortize the cost of learning the channel state information at the receivers (CSIR) and/or the delays in feeding back channel state information to the transmitters (CSIT). Yet, beyond that, it is not known whether \emph{network} coherence\footnote{Network coherence refers to the model where all the channels in the network follow the same coherence pattern, eliminating the diversity of coherence patterns that enables blind interference alignment schemes \cite{Jafar_corr}.}  offers any additional DoF benefits. Specifically, if CSIR is assumed to be perfectly available and the CSIT, limited to finite precision as it may be, is also assumed to be available instantaneously, then it is not known whether the network coherence time still impacts the DoF of interference networks. Partial insights into this question have emerged recently through novel achievable schemes \cite{CBIA, Jafar_TIM, Naderi_Avestimehr}. However, a conclusive answer to this question has remained elusive due to the difficulty of obtaining  DoF outer bounds that are sensitive to network coherence time. In fact, no such bounds exist, to the best of our knowledge. The lack of such bounds is underscored by various open problems noted in \cite{Naderi_Avestimehr, Gou_TIM}.

A promising development in this regard is the recent emergence of an outer bound argument in \cite{Arash_Jafar_PN} based on bounding the cardinality of the images of codewords that align at one receiver but remain distinguishable at another receiver (in short, the Aligned Image Sets (AIS) argument). Motivated by this promising development, in this work we use a novel adaptation of  the AIS approach to prove that indeed network coherence time matters, even with perfect CSIR and instantaneous finite precision CSIT. As immediate application of our result, we are able to settle  the open problems from \cite{Naderi_Avestimehr, Gou_TIM}.

Coherence times are critical for acquiring CSIR or CSIT, as shown in \cite{Hassibi_Hochwald, Hochwald_Marzetta, Lapidoth, Tse_Hanly2, Zheng_Tse}. Even with perfect CSIR and no CSIT except the knowledge of the coherence patterns, the idea of blind interference alignment was introduced in \cite{Jafar_corr} to show that a \emph{diversity} of coherence patterns enables DoF improvements. Blind interference alignment is not feasible if there is no diversity of coherence patterns, i.e., coherence patterns are identical across users (\emph{network} coherence). In this setting, are there further DoF benefits of channel coherence? The recent body of work on topological interference management  \cite{CBIA, Jafar_corr, Naderi_Avestimehr} suggests that there is such a possibility. Introduced in \cite{Jafar_corr}, topological interference management (TIM) refers to DoF studies of partially connected wireless networks with perfect CSIR and no CSIT beyond the network connectivity. As shown in \cite{Jafar_corr}, TIM is essentially related to the index coding problem,  interference alignment plays a crucial part in TIM (and index coding), and DoF gains from interference alignment are achieved even though no knowledge of channel realizations is available to the transmitters provided that the network coherence times are sufficiently long. Reference \cite{CBIA} provides the first example where such gains are achievable even with network coherence time of unity. TIM for unit coherence time $T_c=1$ is then studied extensively  in \cite{Naderi_Avestimehr} by Naderi and Avestimehr, who obtain broad characterizations of the DoF gains possible in this setting. Remarkably, with $T_c=1$, the DoF achieved  in \cite{Naderi_Avestimehr} are in general strictly smaller than what is achieved, say for $T_c=2$ in \cite{Jafar_corr}. Thus, the achievable schemes suggest that coherence time matters. However, in all instances where higher DoF are achieved with a longer coherence time, the optimality of the achievable schemes for the shorter coherence times remains unknown. This is because the outer bounds in \cite{Naderi_Avestimehr} are not sensitive to network coherence times, and thus cannot distinguish between $T_c=1$ and $T_c>1$.  Indeed, to our knowledge no such DoF outer bounds exist anywhere that are sensitive to network coherence times (when CSIR is perfect and CSIT is available without delay). In this paper we present the first such outer bound, based on the Aligned Image Sets approach \cite{Arash_Jafar_PN}. The new bound proves that indeed network coherence time matters for interference networks with perfect CSIR and finite precision CSIT. It also allows us to settle open problems previously noted in \cite{Naderi_Avestimehr,Gou_TIM}. Two open problems where a gap remains between the achievable DoF of \cite{Naderi_Avestimehr} and the DoF outer bounds of \cite{Naderi_Avestimehr} are highlighted by Naderi and Avestimehr (cf. Figure 16 of \cite{Naderi_Avestimehr}). The problems  are reproduced in this paper in Figure \ref{Fig2}. Optimal DoF for both problems are immediately settled by the new outer bound derived in this paper. A related open problem is the achievability of $1/2$ DoF per user in the TIM setting with coherence time $T_c=1$. In \cite{Gou_TIM}, Gou et al. characterize a sufficient condition for achievability of $1/2$ DoF per user,  However, in the absence of an outer bound for the $T_c=1$ setting, it remains unknown whether the sufficient condition of Gou et al. is also a necessary condition. Our new outer bound also settles this open problem, establishing a necessary and sufficient condition for achievability of $1/2$ DoF per user in the TIM setting with coherence time $T_c=1$.

An underlying theme from this and other recent works that successfully generalize the AIS approach in various directions \cite{Arash_Jafar_PN, Arash_Jafar_TC, Arash_Jafar_IC, Arash_Bofeng_Jafar_BC,Arash_Jafar_sumset,Arash_Jafar_MIMOsym_ArXiv}, is the broadening scope of the aligned image sets argument. Recognized  in \cite{Korner_Marton} by Korner and Marton  more than $40$ years ago, characterizing the difference in the size of image sets at different receivers is one of the most essential challenges in network information theory. Seen in this light, interference alignment schemes address this challenge from the achievability side, showing how under various specialized assumptions it is possible to create a large difference, i.e., create a large image at one receiver while the image at the other receiver remains small because of interference alignment. As noted in \cite{Arash_Jafar_PN}, the AIS argument is the  other side of the same coin. It shows, from the converse side, how under various limitations on the precision of CSIT, the difference in the sizes of images cannot be made too large. Indeed, just as interference alignment in its various forms seems inevitable in understanding optimal achievable schemes for wireless networks,  so too the aligned image sets bounds may be equally unavoidable for robust converse arguments. 

\section{Definitions}\label{sec:def}
\begin{figure}
\centerline{\includegraphics[width=5.2in]{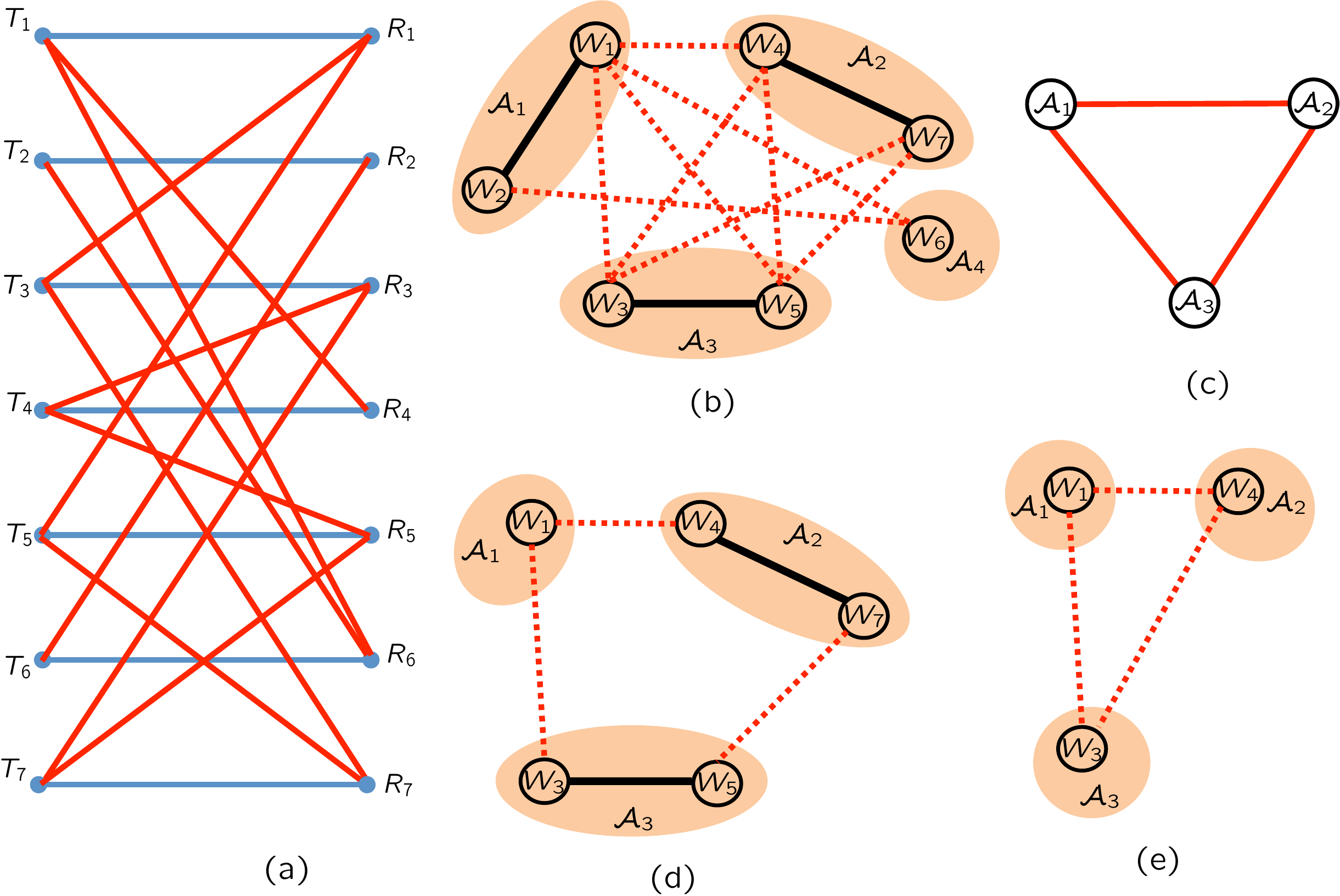}}
\caption{(a) Partially connected interference network. (b) Corresponding Alignment graph (black edges) and Conflict graph (dashed red edges). Also shown are the alignment sets $\mathcal{A}_1, \mathcal{A}_2,\mathcal{A}_3, \mathcal{A}_4$. (c) Reduced graph $\mathbb{G}_r$ comprised of $\mathcal{A}_1, \mathcal{A}_2,\mathcal{A}_3$. Note that $\mathcal{A}_4$ is not a part of $\mathbb{G}_r$ because it has only one message. Also note that $\mathbb{G}_r$ has an odd cycle $\mathcal{C}_r$ of length $m=3$. (d) A completed cycle corresponding to $\mathcal{C}_r$, for which $m=3, m_2=1, l_\Sigma=3$. } \label{fig:graphs}
\end{figure}
The following definitions of undirected graphs originate in the topological interference management framework of \cite{Jafar_TIM}.\begin{definition}[Alignment Graph $\mathbb{G}_a$ and Alignment Set $\mathcal{A}_s$] The vertices of the alignment graph are the $K$ messages, $W_1, W_2, \cdots, W_K$. Messages $W_i$ and $W_j$ are connected with a solid black edge (called an alignment edge) if the sources of both these messages are heard by a destination that desires message $W_k\notin\{W_i, W_j\}$. Each connected component of the alignment graph is called an alignment set.
\end{definition}

\begin{definition}[Conflict Graph $\mathbb{G}_c$ and Internal Conflict] The vertices of the conflict graph are  the $K$ messages, $W_1,W_2,\cdots,W_K$. Message $W_i$ is connected by a dashed red edge (called a conflict edge) to  all other messages $W_j$ whose sources are heard by the destination that desires message $W_i$.  If two messages that belong to the same alignment set have a conflict edge between them, it is called an internal conflict.
\end{definition}

\begin{definition}[Reduced Graph $\mathbb{G}_r$] The vertices of the reduced graph $\mathbb{G}_r$ are those alignment sets $\mathcal{A}_i$ that have two or more messages, i.e., $|\mathcal{A}_i|\geq 2$. Singleton alignment sets are not represented in $\mathbb{G}_r$.  $\mathcal{A}_i$ and $\mathcal{A}_j$ in $\mathbb{G}_r$ have an edge between them if the conflict graph contains an edge between a message  $W_i\in\mathcal{A}_i$ and a message $W_j\in\mathcal{A}_j$.
\end{definition}

\begin{definition}[Completed Cycle $\mathcal{C}_c$ and parameters $m, m_2, l_\Sigma$]\label{deflsigm} A completed cycle is a relation from a cycle in $\mathbb{G}_r$ to a cycle in another graph where the vertices are the messages and each edge is either an alignment edge or a conflict edge. It is obtained as follows. Consider a cycle $\mathcal{C}_r$  in $\mathbb{G}_r$, of length $m$, that is comprised of edges $(\mathcal{A}_{i_1}, \mathcal{A}_{i_2})$, $(\mathcal{A}_{i_2},\mathcal{A}_{i_3})$, $\cdots$, $(\mathcal{A}_{i_{m-1}}, \mathcal{A}_{i_m})$, $(\mathcal{A}_{i_m},\mathcal{A}_{i_1})$. A  completed cycle $\mathcal{C}_{c}$ that is related to $\mathcal{C}_r$ is obtained by replacing each edge $(\mathcal{A}_{i_j},\mathcal{A}_{i_{j+1}})$ of $\mathcal{C}_r$ (subscripts interpreted cyclically, so that $i_{m+1}=i_1$) with a conflict edge $(W_{i_j}, W_{i_{j+1}}')$, $W_{i_j}\in\mathcal{A}_{i_j}$, $W_{i_{j+1}}'\in\mathcal{A}_{i_{j+1}}$. Each vertex $\mathcal{A}_{i_j}$  of $\mathcal{C}_r$ is replaced with the message $W_{i_j}$ if $W_{i_j}=W_{i_j}'$, or by a path from $W_{i_j}$ to $W_{i_j}'$ comprised of alignment edges connecting a subset of messages drawn from $\mathcal{A}_{i_j}$ if $W_{i_j}\neq W_{i_j}'$. The resulting graph is a cycle, called completed cycle, which contains exactly $m$ conflict edges. All the remaining edges are alignment edges. Define $m_2$ as the number of  instances of ${i_j}\in\{1,2,\cdots,m\}$ for which $W_{i_j} = W_{i_j}'$. Further, if the length of the completed cycle is denoted as $|\mathcal{C}_c|$, then define $l_\Sigma\triangleq |\mathcal{C}_c|-m+m_2$.
\end{definition}

The next three definitions are related to the finite precision channel knowledge assumption.

\begin{definition}[Bounded Density Channel Coefficients] Define a set of real valued random variables, $\mathcal{G}$ such that the magnitude of each random variable $g\in\mathcal{G}$ is bounded away from zero and infinity, $0<\Delta_1\leq |g|\leq\Delta_2<\infty$, for some constants $\Delta_1,\Delta_2$, and there exists a finite positive constant $f_{\max}$, such that for all finite cardinality disjoint subsets $\mathcal{G}_1, \mathcal{G}_2$ of $\mathcal{G}$, the joint probability density function of all random variables in $\mathcal{G}_1$, conditioned on all random variables in $\mathcal{G}_2$, exists and is bounded above by $f_{\max}^{|\mathcal{G}_1|}$. Without loss of generality we will assume that $f_{\max}\geq 1, \Delta_2\geq 1$.
\end{definition}

\begin{definition}[Arbitrary Channel Coefficients] Let $\mathcal{H}$ be a set of arbitrary constant values that are bounded above by $\Delta_2$, i.e., if $h\in\mathcal{H}$ then $|h|\leq\Delta_2<\infty$.
\end{definition}

\begin {definition}[Bounded Density Linear Combinations]For real numbers $x_1,x_2,\cdots,x_k$ define the notations $L_j^{b}(x_i,1\le i\le k)$, and $L_j(x_i,1\le i\le k)$ to represent,
\begin {eqnarray}
L^{b}_j(x_1,\cdots,x_k )\triangleq\sum_{1\le i\le k} \lfloor g_{j_i}x_i\rfloor\\
L_j(x_1,\cdots,x_k)\triangleq\sum_{1\le i\le k} \lfloor h_{j_i}x_i\rfloor
\end{eqnarray}
for  ~distinct ~random~ variables ~$g_{j_i}\in\mathcal{G}$, ~and~ for  ~arbitrary  constants $h_{j_i}\in\mathcal{H}$. The corresponding multi-letter forms are defined as 
\scalebox{0.81}{%
$L^{b[n]}_j(x_1,\cdots,x_k )\triangleq \left(\sum_{1\le i\le k} \lfloor g_{j_i}(1)x_i(1)\rfloor, \cdots, \sum_{1\le i\le k} \lfloor g_{j_i}(n)x_i(n)\rfloor\right),$
}
\scalebox{0.81}{%
$L^{[n]}_j(x_1,\cdots,x_k )\triangleq \left(\sum_{1\le i\le k} \lfloor h_{j_i}(1)x_i(1)\rfloor, \cdots, \sum_{1\le i\le k} \lfloor h_{j_i}(n)x_i(n)\rfloor\right),$
}
 for distinct $g_{j_i}(t)\in\mathcal{G}$ and arbitrary constants $h_{j_i}\in\mathcal{H}$. We refer to the $L^{b}$ functions as  bounded density linear combinations.
\end {definition}
Finally, for compact notation, let us define $[k]=\{1,2,\cdots, k\}$ for positive integer $k$.

\section{System Model} {\label{sec-sys}}
\subsection{The Channel}
Under the DoF framework, the channel model for the partially connected\footnote{A DoF characterization for the partially connected setting is a special case of the GDoF characterization for arbitrary channel strength levels. As such, the main insights are not limited to binary connectivity models, i.e., the DoF gap due to coherence time for partially connected channels can be readily translated into a GDoF gap due to coherence time for channels with sufficiently disparate strengths.} $K$ user interference channel is defined by the following input-output equations. $\forall k\in[K]$,
\begin{eqnarray}
Y_k(t)&=&\sqrt{P}G_{kk}(t)X_k(t)+\sum_{l\in \mathcal{M}_k}\sqrt{P}G_{kl}(t)X_l(t)+Z_k(t).\label{chm}
\end{eqnarray}
The channel uses are indexed by $t\in\mathbb{N}$, $X_l(t)$ is the symbol sent from transmit antenna $l$ subject to a  unit power constraint, $Y_k(t)$ is the symbol observed by Receiver $k$, $Z_k(t)$ is the zero mean unit variance additive white Gaussian noise (AWGN) at Receiver $k$, and $G_{kl}(t)$ is the channel fading coefficient between Transmitter $l$ and Receiver $k$. We assume perfect channel state information at the receivers (CSIR), but the channel state information at the transmitters (CSIT) is limited to finite precision, i.e., $\forall k\in[K], l\in[K], t\in\mathbb{N}$, $G_{kl}(t)$ are distinct elements of $\mathcal{G}$. Note that this implies that the coherence time $T_c=1$.\footnote{While the channel coefficients change with every channel use,  note that we do not require that they should be \emph{independent} across $t$. Our results hold whether the channels take independent values or remain correlated in time, provided the joint density functions are bounded.} The transmitters are  aware of the joint probability density function (pdf) of the channel coefficients, which satisfies the bounded density assumption. Beyond this, the transmitters have no knowledge of the channel realizations. Thus, the transmitted symbols $X_l(t)$ may  depend on the pdf of $\mathcal{G}$ but are independent of the realizations of $\mathcal{G}$. 
$P$ is the nominal SNR parameter that is allowed to approach infinity. The partial connectivity is specified through the set $\mathcal{M}_k$ which is  defined as a subset of the set $[K]$, such that $l\in \mathcal{M}_k$ if and only if the $l$-th transmitter can be heard by the $k$-th receiver. For simplicity, let us assume all values are real. Generalizations to complex channels are somewhat cumbersome but  conceptually straightforward as in \cite{Arash_Jafar_PN}.

\subsection{Finite Precision CSIT}
Under finite precision CSIT, the channel coefficients may be represented as
\begin{eqnarray}
G_{kl}(t)&=&\hat{G}_{kl}(t)+\tilde{G}_{kl}(t)
\end{eqnarray}
Recall that for any $k,l\in[K]$, $G_{kl}(t)$ is the channel fading coefficient between Transmitter $l$ and Receiver $k$. $\hat{G}_{kl}(t)$ are the  channel estimate terms and $\tilde{G}_{kl}(t)$ are the  estimation error terms. To avoid degenerate conditions, the ranges of values are bounded away from zero and infinity as follows, i.e., there exist constants $\Delta_1, \Delta_2$ such that $0<\Delta_1\leq 
|{G}_{kl}(t)|$, and $|\hat{G}_{kl}(t)|,|\tilde{G}_{kl}(t)|<\Delta_2<\infty$. The channel variables $\hat{G}_{kl}(t), \tilde{G}_{kl}(t)$, $\forall k,l\in[K], t\in\mathbb{N}$, are subject to the bounded density assumption with the difference that the actual realizations of $\hat{G}_{kl}(t)$ are revealed to the transmitter, but the realizations of $\tilde{G}_{kl}(t)$ are not available to the transmitter.
\subsection{DoF}
The definitions of achievable rates $R_i(P)$ and capacity region $\mathcal{C}(P)$ are standard. The DoF region is defined as
\begin{eqnarray}
\mathcal{D}&=&\{(d_1,\cdots,d_K): \exists (R_1(P),\cdots, R_K(P))\nonumber\\
&&\in\mathcal{C}(P), \mbox{ s.t. } d_k=\lim_{P\rightarrow\infty}\frac{R_k(P)}{\frac{1}{2}\log(P)}, \forall k\in[K]\} 
\end{eqnarray}

\section{Results: Coherence Time Matters}\label{sec:mainresult}
The main contribution of this work is an outer bound, based on the aligned images argument, which shows that the DoF of an interference network under finite precision CSIT and perfect CSIR, are limited by the network coherence time, i.e., coherence time matters. In particular, we bound the DoF under coherence time $T_c=1$ and show that this bound is strictly smaller than what is achievable in general with a larger coherence time, say $T_c=2$.
\begin{theorem}\label{theorem:main}
For a partially connected $K$ user interference channel with finite precision CSIT and coherence time $T_c=1$, if the reduced graph $\mathbb{G}_r$ has an odd-length cycle $\mathcal{C}_r$,then the following bound holds on the symmetric DoF per user.
\begin{eqnarray}
\mbox{Symmetric DoF per User, }\alpha&{\leq}&\left(\frac{1}{2}\right)\left(1-\frac{1}{m+2l_\Sigma}\right)\label{eq:main}
\end{eqnarray}
where the parameters $m, m_2$ and $l_\Sigma$ are as defined in Section \ref{sec:def} for any completed cycle $\mathcal{C}_c$ related to $\mathcal{C}_r$.
\end{theorem}
For the interference network illustrated in Figure \ref{fig:graphs}(a), the reduced graph $\mathbb{G}_r$ (shown in Figure \ref{fig:graphs}(c) has cycle of odd length $m=3$. The completed graph in Figure \ref{fig:graphs}(d) has $m=3, m_2=1, l_\Sigma=3$, so the outer bound (\ref{eq:main}) from Theorem \ref{theorem:main} tells us that symmetric DoF per user $\leq 4/9$.   In fact $4/9$ is achievable, see Section \ref{4/9ach}.

As an immediate application of Theorem \ref{theorem:main}, we have the following corollary which settles an open problem from \cite{Gou_TIM}.

\begin{corollary}\label{corollary:half}
In a partially connected $K$ user interference channel with finite precision CSIT and coherence time $T_c=1$, the symmetric DoF value of $1/2$ per user is achievable if and only if the following two conditions are satisfied.
\begin{enumerate}
\item[C1.] There are no internal conflicts.
\item[C2.] The reduced graph $\mathbb{G}_r$ has no odd length cycles.
\end{enumerate}
\end{corollary}
\proof The achievability result, i.e., that conditions {\it C1}, {\it C2} are sufficient for achieving a symmetric DoF of $1/2$ per user, was  established by Gou et al. (Theorem 1 in \cite{Gou_TIM}) utilizing the topological interference management framework of  \cite{Jafar_TIM}. Gou et al. assume that the transmitters are not aware of the coherence time, and show that $1/2$ DoF per user is achievable regardless of the length of the coherence interval when conditions {\it C1}, {\it C2} are satisfied.  The necessity of {\it C1} is established in \cite{Jafar_TIM}, which shows that if there are internal conflicts then the symmetric DoF per user are strictly less than $1/2$. This is shown for arbitrarily large coherence times, so it holds for coherence time $T_c=1$ as well. The necessity of Condition {\it C2} was previously open but is immediately settled by Theorem \ref{theorem:main}, because the presence of an odd cycle in $\mathbb{G}_r$ activates the outer bound (\ref{eq:main}) which means that the symmetric DoF value per user is strictly less than $1/2$. \hfill\QED

Note that the result of Corollary \ref{corollary:half} holds even if the transmitters are unaware of the value of the coherence time. This is because an achievable scheme that works for all coherence times, must also work for coherence time $T_c=1$.

\begin{figure}[!t]
\centerline{\includegraphics[height=2.7in]{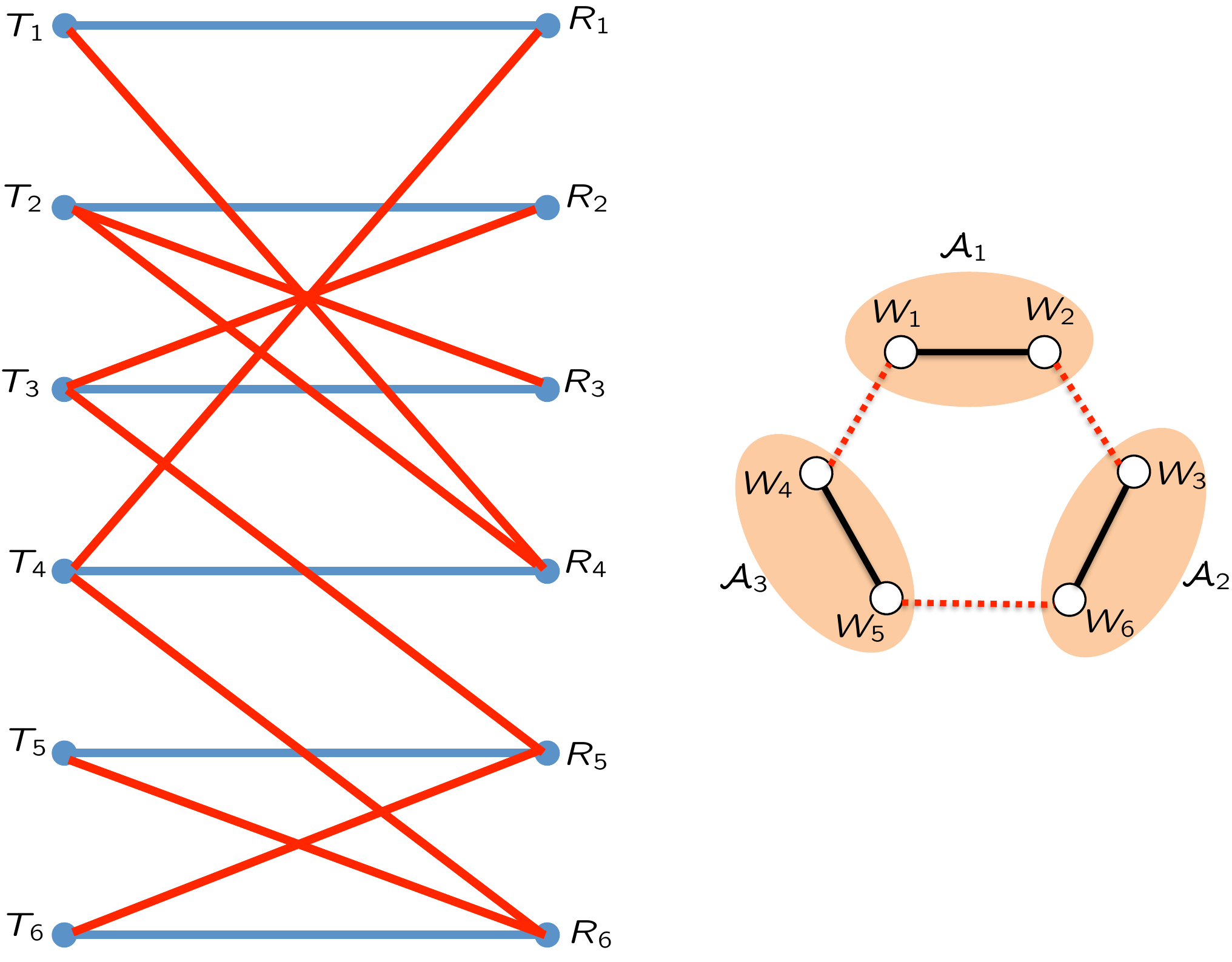}}
\caption{First open problem from \cite{Naderi_Avestimehr} (see Figure 16 of \cite{Naderi_Avestimehr}).}\label{Fig2}
\end{figure}

\begin{figure}[!t]
\centerline{\includegraphics[height=2.7in]{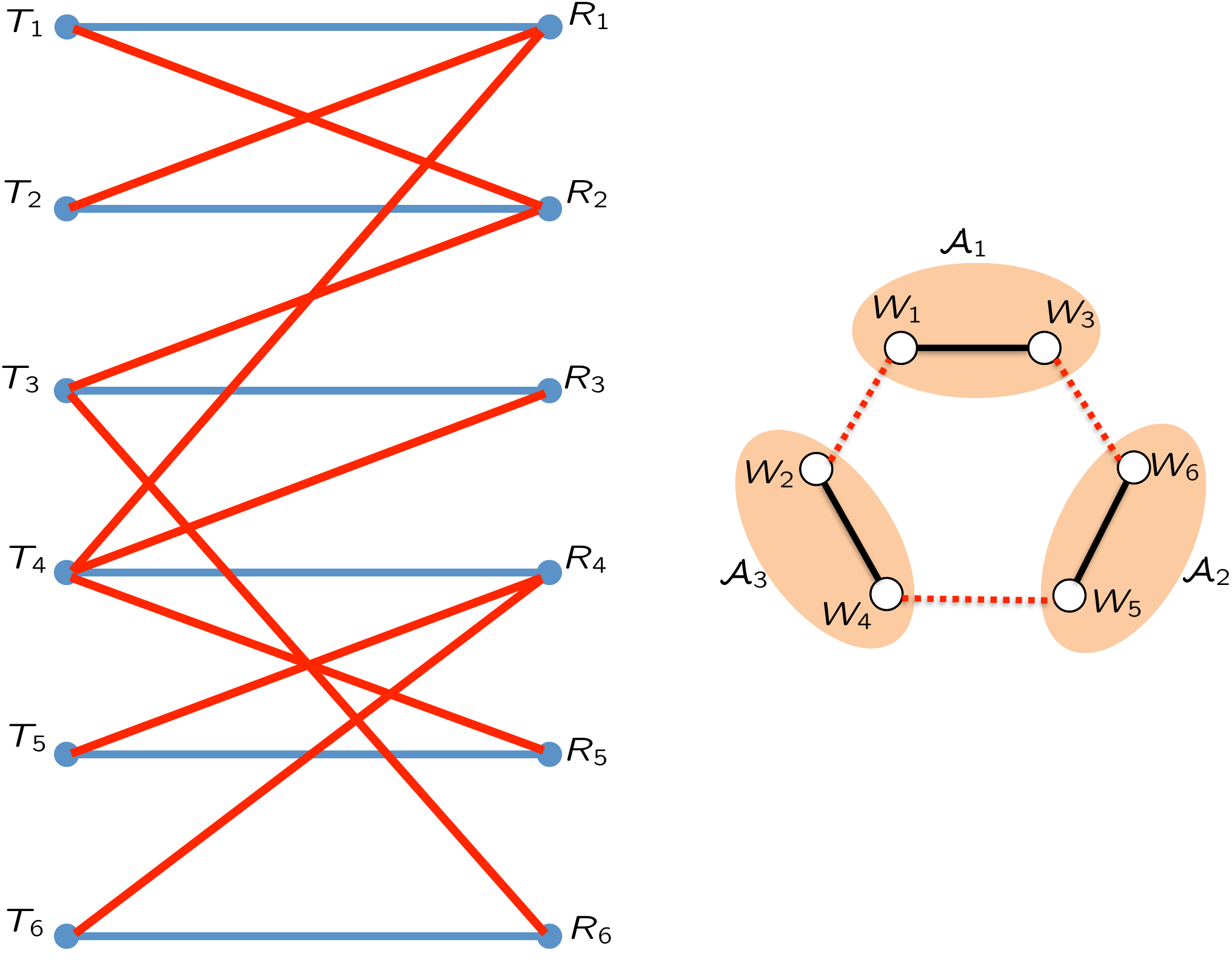}}
\caption{Second open problem from \cite{Naderi_Avestimehr} (see Figure 16 of \cite{Naderi_Avestimehr}).}\label{Fig3}
\end{figure}

As another application of the new bound,  consider  the two examples of open problems highlighted by Naderi and Avestimehr in \cite{Naderi_Avestimehr} (see Figure 16 of \cite{Naderi_Avestimehr}) where the optimal symmetric DoF per user are unknown for $T_c=1$. The two examples are illustrated in Figure  \ref{Fig2} and Figure \ref{Fig3}.

References  \cite{Gou_TIM} and \cite{Naderi_Avestimehr} have shown that the $\alpha=4/9$ is achievable in each of these settings. However, the best outer bound previously known is $\alpha\leq 1/2$, which is achievable (and optimal) if coherence time is greater than or equal to $2$, as shown in \cite{Jafar_TIM}. A tight outer bound was not previously available when coherence time is unity. However, the following corollary of Theorem \ref{theorem:main} settles the symmetric DoF per user for coherence time $T_c=1$ for both of these networks.

\begin{corollary}\label{corollary:Naderi}
For each of the partially connected interference networks illustrated in Figure \ref{Fig2}, with coherence time $T_c=1$, the optimal symmetric DoF per user $= 4/9$.
\end{corollary}
\proof For each of the networks, from the cycles of reduced graph illustrated in Figure \ref{Fig2}, we have $m=3, m_2=0$ and $l_\Sigma=3$.  Substituting into  (\ref{eq:main}) we find the outer bounds $\alpha\leq 4/9$, thus settling the symmetric DoF for both of these networks.\hfill\QED

\section{Proof of Theorem \ref{theorem:main}}
\begin{figure}[!h]
\centerline{\includegraphics[width=4.1in]{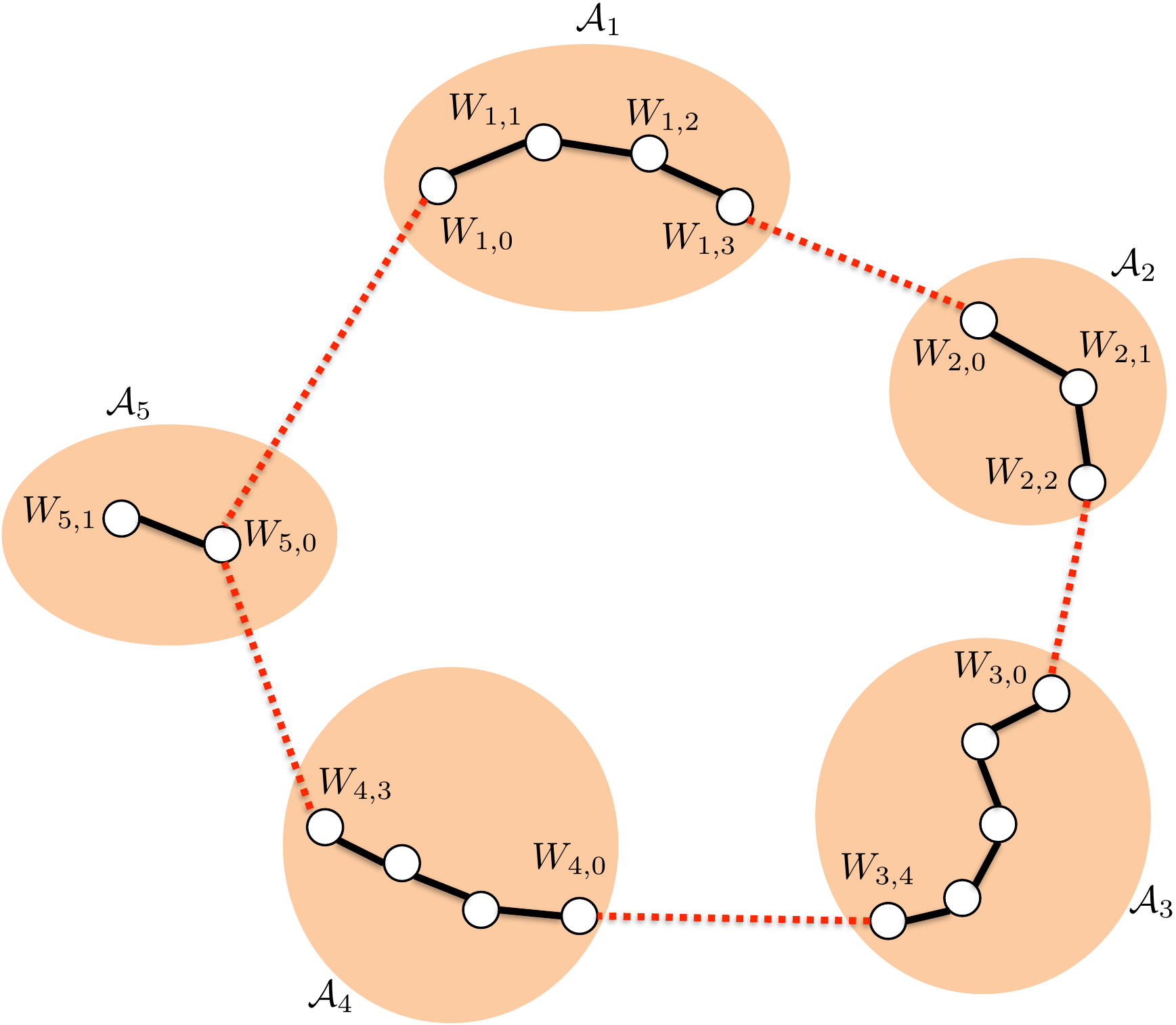}}
\caption{Completed cycle with $m=3, m_2=1, l_\Sigma=13$. }\label{Fig1}
\end{figure}

Suppose there exists a cycle of odd length $m$ in the reduced graph $\mathbb{G}_r$. Then there exist alignment sets $\mathcal{A}_1,\mathcal{A}_2, \cdots, \mathcal{A}_{m}$, such that there exists a conflict between any two consecutive sets, $\mathcal{A}_i, \mathcal{A}_{i+1}$. Note that the indices are interpreted in a cyclic manner, so that $\mathcal{A}_1$ follows $\mathcal{A}_m$. Consider alignment set $\mathcal{A}_i$. Choose a message $W_i\in\mathcal{A}_{i}$ such that $W_i$ conflicts with a message in $\mathcal{A}_{i-1}$. Similarly, choose a message  $W_i'\in\mathcal{A}_{i}$ that conflicts with a message in $\mathcal{A}_{i+1}$. If $W_i\neq W_i'$, then find the shortest path from $W_i$ to $W_i'$, comprised of alignment edges. Such a path exists because  $W_i,W_i'\in\mathcal{A}_i$ and $\mathcal{A}_i$ is a connected component of the alignment graph. Let the length of this path be $l_i$. Without loss of generality, label the messages along this path as $W_i=W_{i,0}, W_{i,1},\cdots, W_{i,l_i}=W_i'$. If $W_i=W_i'$, then choose a different message $W_i''\in\mathcal{A}_i$ which is connected to $W_i$ with an alignment edge. Such a message must exist because  each alignment set involved in the reduced graph has two or more messages. In this case, the path from $W_i$ to $W_i''$ is of length $l_i=1$, and without loss of generality we label $W_i=W_{i,0}, W_i''=W_{i,l_i}$. Such a situation occurs in $\mathcal{A}_5$ in the example illustrated in Figure \ref{Fig1}. Other messages and conflict/alignment edges may exist, but are not important for this proof, so they are suppressed for clarity in Figure \ref{Fig1}.  Define
\begin{eqnarray}
l'_i&\triangleq&\left\{
\begin{array}{ll}
l_i&\mbox{ if } W_i\neq W_i'\\
0&\mbox{ if }  W_i= W_i'
\end{array}
\right.
\end{eqnarray}

\subsection{Alignments $Z_{\checkmark}^{b }$ and Conflicts $Z_{\times}^{b }$}
Following in the steps of the AIS argument of \cite{Arash_Jafar_PN}, we use the deterministic approximation of (\ref{chm}) with integer-valued inputs $\bar{X}_k(t)\in\{0,1,\cdots,\bar{P}\}$ and  integer-valued outputs $\bar{{Y}}_{k}(t), k\in[K]$, so that 
\begin{eqnarray}
\bar{Y}_k(t)&=&\left \lfloor G_{kk}(t)\bar{X}_k(t)\right \rfloor+\sum_{l\in \mathcal{M}_k}\left \lfloor G_{kl}(t)\bar{X}_l(t)\right \rfloor \label{chm2}
\end{eqnarray}
and $\bar{P}$ is defined as  $\left \lfloor \sqrt{P}\right \rfloor$. For ease of exposition, let us further customize our notation for the completed cycle. For the transmitter sending message $W_{i,j}$, denote the transmitted symbols as $\bar{X}_{i,j}$. Further, define $Z_{\checkmark}^{b}$ and $Z_{\times}^{b}$ as follows. The time index is suppressed for compact notation.
\begin{eqnarray}
Z^{b}_{\checkmark}&=&(L^{b}_{1{\checkmark}}(\bar{X}_{1,0},\bar{X}_{1,l_1}),L^{b}_{2{\checkmark}}(\bar{X}_{2,0},\bar{X}_{2,l_2}),\cdots,L^{b}_{m{\checkmark}}(\bar{X}_{m,0},\bar{X}_{m,l_{m}})), \\
Z^{b}_{\times}&=&(L^{b}_{1\times}(\bar{X}_{1,{l'_1}},\bar{X}_{2,0}),L^{b}_{2\times}(\bar{X}_{2,l'_2},\bar{X}_{3,0}),\cdots,L^{b}_{m\times}(\bar{X}_{m,l'_{m}},\bar{X}_{1,0})).
\end{eqnarray}
Note that we used $l_i$ in the term $Z^{b}_{\checkmark}$ and $l'_i$ in the term $Z^{b}_{\times}$. For the example illustrated in Figure \ref{Fig1} these would be
\begin{eqnarray}
Z^{b}_{\checkmark}&=&(L^{b}_{1{\checkmark}}(\bar{X}_{1,0}^{},\bar{X}_{1,3}^{}),L^{b}_{2{\checkmark}}(\bar{X}_{2,0}^{},\bar{X}_{2,2}^{}),L^{b}_{3{\checkmark}}(\bar{X}_{3,0}^{},\bar{X}_{3,4}^{}),L^{b}_{4{\checkmark}}(\bar{X}_{4,0}^{},\bar{X}_{4,3}^{}),{\color{black}L^{b}_{5{\checkmark}}(\bar{X}_{5,0}^{},\bar{X}_{5,1}^{})}) \nonumber\\
Z^{b}_{\times}&=&(L^{b}_{1\times}(\bar{X}_{1,3}^{},\bar{X}_{2,0^{}}),L^{b}_{2\times}(\bar{X}_{2,2}^{},\bar{X}_{3,0}^{}),L^{b}_{3\times}(\bar{X}_{3,4}^{},\bar{X}_{4,0}^{}),L^{b}_{4\times}(\bar{X}_{4,3}^{},\bar{X}_{5,0}^{}),L^{b}_{5\times}({\color{black}\bar{X}_{5,0}}^{},\bar{X}_{1,0}^{})).\nonumber\\
\end{eqnarray}
Multi-letter forms, $Z_\checkmark^{b[n]}, Z_\times^{b[n]}$ are obtained by replacing  $L_{i\checkmark}^{b}, L_{i\times}^{b}$ with $L_{i\checkmark}^{b[n]}, L_{i\times}^{b[n]}$, respectively. The intuitive significance of the notation is as follows. We use $\checkmark$  as a subscript for  combinations of symbols that we would like to align because these are messages connected by alignment edges, while $\times$ is used as a subscript for  combinations of symbols that we would like to \emph{not} align, because of message conflicts.

The symmetric DoF bound that we seek will come from bounding $H(Z^{b[n]}_{\times}|\mathcal{G})-H(Z^{b[n]}_{\checkmark}|\mathcal{G})$ from above and from below. Let us start with the lower bound.

\subsection{Bounding $H(Z^{b[n]}_{\times}|\mathcal{G})-H(Z^{b[n]}_{\checkmark}|\mathcal{G})$ from below}
In order to derive a lower bound on  $H(Z^{b[n]}_{\times}|\mathcal{G})-H(Z^{b[n]}_{\checkmark}|\mathcal{G})$, we will derive an upper bound on the negative term $H(Z^{b[n]}_{\checkmark}|\mathcal{G})$  and a lower bound on the positive term $H(Z^{b[n]}_{\times}|\mathcal{G})$. These bounds are based on alignment and conflict graphs, i.e., the topological interference management perspective.
\subsubsection{Bounding $H(Z^{b[n]}_{\checkmark}|\mathcal{G})$ from above}
Let us first bound the terms $H(L_{i\checkmark}^{b[n]}(\bar{X}_{i,0},\bar{X}_{i,l_i})|\mathcal{G})$. Note that $\forall j\in\{0,1,\cdots, l_i-1\}$,
\begin{eqnarray}
H(L_{i\checkmark}^{b[n]}(\bar{X}_{i,j},\bar{X}_{i,j+1})|\mathcal{G})&\leq&(1-\alpha)n\log(\bar{P})
\end{eqnarray}
This is because $W_{i,j}, W_{i,j+1}$ are connected by an alignment edge, i.e., both messages cause interference at a receiver where neither is desired. Since $\alpha$ dimensions must be left interference free for the desired message, the collective interference at this receiver from $W_{i,j}, W_{i,j+1}$, i.e., $H(L_{i\checkmark}^{b[n]}(\bar{X}_{i,j},\bar{X}_{i,j+1})|\mathcal{G})$ must have no more than $(1-\alpha)$ DoF.

Further, using the functional form of submodularity property of the entropy function for arbitrary random variables $U_1, U_2, U_3$,
\begin{eqnarray}
H(U_1,U_2,U_3)+H(U_1+U_2+U_3)&\leq&H(U_1+U_2,U_3)+H(U_1,U_2+U_3)
\end{eqnarray}
and for independent $U_1, U_2, U_3$, 
\begin{eqnarray}
H(U_2)+H(U_1+U_2+U_3)&\leq&H(U_1+U_2)+H(U_2+U_3)
\end{eqnarray}
let us proceed as follows (as usual, $o(\log(P))$ terms that are inconsequential for DoF are suppressed),
\begin{eqnarray}
H(L_{1\checkmark}^{b[n]}(\bar{X}_{i,0},\bar{X}_{i,1})|\mathcal{G})
&\leq&n(1-\alpha)\log(\bar{P})\\
H(L_{1\checkmark}^{b[n]}(\bar{X}_{i,0},\bar{X}_{i,2})|\mathcal{G})
&\leq&H(L_{1\checkmark}^{b[n]}(\bar{X}_{i,0},\bar{X}_{i,1})|\mathcal{G})+H(L_{1\checkmark}^{b[n]}(\bar{X}_{i,1},\bar{X}_{i,2})|\mathcal{G})-H(\bar{X}_{i,1}^{[n]})\nonumber\\
&\leq&\Big(2(1-\alpha)-\alpha\Big)n\log(\bar{P})\\
H(L_{1\checkmark}^{b[n]}(\bar{X}_{i,0},\bar{X}_{i,3})|\mathcal{G})&\leq&H(L_{1\checkmark}^{b[n]}(\bar{X}_{i,0},\bar{X}_{i,2})|\mathcal{G})+H(L_{1\checkmark}^{b[n]}(\bar{X}_{i,2},\bar{X}_{i,3})|\mathcal{G})-H(\bar{X}_{i,2}^{[n]})\nonumber\\
&\leq&\Big(3(1-\alpha)-2\alpha\Big)n\log(\bar{P})\\
\vdots&&\nonumber\\
H(L_{1\checkmark}^{b[n]}(\bar{X}_{i,0},\bar{X}_{i,l_i})|\mathcal{G})
&\leq&\Big(l_i(1-\alpha)-(l_i-1)\alpha\Big)n\log(\bar{P})\nonumber
\end{eqnarray}
Finally, because $X_{i,j}$ are all independent, we have the bound,
\begin{eqnarray}
H(Z^{b[n]}_{\checkmark}|\mathcal{G})&=&\sum_{i=1}^{m}H(L_{i\checkmark}^{b[n]}(\bar{X}_{i,0},\bar{X}_{i,l_i})|\mathcal{G})\\
&\leq&\Big(l_\Sigma(1-2\alpha)+m\alpha\Big)n\log(\bar{P})
\end{eqnarray}
where $l_\Sigma\triangleq l_1+l_2+\cdots+l_{m}=\sum_{i=1}^{m}l'_i+m_2$.

\subsubsection{Bounding $H(Z^{b[n]}_{\times}|\mathcal{G})$ from below}
For this, we need to bound the terms $H(L_{i\times}^{b[n]}(\bar{X}_{i,{\color{black}l'_i}},\bar{X}_{i+1,0})|\mathcal{G})$. Recall that the messages were chosen such that  $W_i'=W_{i,{\color{black}l'_i}}$ conflicts with $W_{i+1}=W_{i+1,0}$. Since conflicting messages cannot align, we must have
\begin{eqnarray}
H(L_{i\times}^{b[n]}(\bar{X}_{i,{\color{black}l'_i}},\bar{X}_{i+1,0})|\mathcal{G})&\geq&2\alpha n\log(\bar{P})
\end{eqnarray}

Finally, because $X_{i,j}$ are all independent, we have the bound,
\begin{eqnarray}
H(Z^{b[n]}_{\times}|\mathcal{G})&=&\sum_{i=1}^{m}H(L_{i\times}^{b[n]}(\bar{X}_{i,l'_i},\bar{X}_{i+1,0})|\mathcal{G})\\
&\geq&2\alpha mn\log(\bar{P})\nonumber
\end{eqnarray}

Combining the bounds obtained for $H(Z^{b[n]}_{\times}|\mathcal{G})$ and $H(Z^{b[n]}_{\checkmark}|\mathcal{G})$, we have
\begin{eqnarray}
&&H(Z^{b[n]}_{\times}|\mathcal{G})-H(Z^{b[n]}_{\checkmark}|\mathcal{G})\nonumber\\
&\geq&\Big(\alpha m+(2\alpha-1)l_\Sigma\Big)\times n\log(\bar{P})\label{eq:general}
\end{eqnarray}
Note that if we set $\alpha=1/2$, then 
\begin{eqnarray}
H(Z^{b[n]}_{\times}|\mathcal{G})-H(Z^{b[n]}_{\checkmark}|\mathcal{G})&\geq&\Big(\frac{m}{2}\Big)n\log(\bar{P})\label{eq:contradict}
\end{eqnarray}

\subsection{Bounding $H(Z^{b[n]}_{\times}|\mathcal{G})-H(Z^{b[n]}_{\checkmark}|\mathcal{G})$ from above: Aligned Image Sets}
This is where the AIS argument is invoked. The steps that are essentially identical to \cite{Arash_Jafar_PN} are summarized here for the sake of completeness. The main novelty appears in the part (\ref{eq:critical})-(\ref{eq:Delta2}).
\begin{eqnarray}
H(Z^{b[n]}_{\times}|\mathcal{G})-H(Z^{b[n]}_{\checkmark}|\mathcal{G})&\leq&\Big(\frac{m-1}{2}\Big)n\log(\bar{P})
\end{eqnarray}
\subsubsection{Replacing $Z^{b}_\times$ with $Z_\times$} While $Z^{b}_\times$ is comprised of bounded density linear combinations, the bound that we derive in this section will be shown in a stronger sense, i.e., it holds for arbitrary linear combinations. So we will bound $H(Z_{\times}^{[n]})-H(Z^{b[n]}_{\checkmark}|\mathcal{G})$ where 
\begin{eqnarray}
Z_{\times}&=&(L_{1\times}(\bar{X}_{1,l'_1},\bar{X}_{2,0}),\cdots,L_{m\times}(\bar{X}_{m,l'_m},\bar{X}_{1,0})).
\end{eqnarray}
$L_{i\times}$ are arbitrary linear combinations, and the codewords $\bar{X}_{i,j}$ are designed with full knowledge of these combinations. Note that $Z^{b[n]}_\checkmark$ remains unchanged, i.e., it is still comprised of bounded density linear combinations $L_{i\checkmark}^{b[n]}$, as before. So the codewords may depend only on the (bounded) probability density functions of the combining coefficients $\mathcal{G}$ but are independent of the actual realizations of the  bounded density combining coefficients.
\subsubsection{Functional Dependence}
There are multiple codewords that may produce  the same $Z_\times^{[n]}$, one of which is chosen according to a random choice function $\mathcal{L}$. Conditioning reduces entropy, so $H(Z^{b[n]}_\checkmark|\mathcal{G})\geq H(Z^{b[n]}_\checkmark|\mathcal{G},\mathcal{L})$, and the minimum over $\mathcal{L}$ (say the minimum corresponds to $\mathcal{L}=\mathcal{L}^*$) is smaller than or equal to the average over $\mathcal{L}$. Our goal is to maximize $H(Z_\times^{[n]})-H(Z^{b[n]}_\checkmark|\mathcal{G})$. Setting $\mathcal{L}=\mathcal{L}^*$ does not change the first term while it can only reduce the second term. Therefore, without loss of generality we will assume henceforth that $\mathcal{L}=\mathcal{L}^*$, i.e., all the codewords $\bar{X}_{i,j}^{[n]}$ are  functions of $Z_\times^{[n]}$. Note that this implies that $Z^{b[n]}_\checkmark$ is a function of $(Z_\times^{[n]},\mathcal{G})$. When needed, for clarity we may highlight this functional dependence by writing $\bar{X}_{i,j}^{[n]}$ as $\bar{X}_{i,j}^{[n]}(Z_\times^{[n]})$ and $Z^{b[n]}_\checkmark$ as $Z^{b[n]}_\checkmark(Z_\times^{[n]},\mathcal{G})$.
\subsubsection{Aligned Image Set}
\begin{eqnarray}
H(Z_\times^{[n]},Z_\checkmark^{b[n]}|\mathcal{G})&=&H(Z_\times^{[n]})+H(Z_\checkmark^{b[n]}|Z_\times^{[n]},\mathcal{G})\\
&=&H(Z_\times^{[n]})\label{eq:fd}\\
H(Z_\times^{[n]},Z_\checkmark^{b[n]}|\mathcal{G})&=&H(Z_\checkmark^{b[n]}|\mathcal{G})+H(Z_\times^{[n]}|Z_\checkmark^{b[n]},\mathcal{G})\\
\implies H(Z_\times^{[n]})-H(Z_\checkmark^{b[n]}|\mathcal{G})&=&H(Z_\times^{[n]}|Z_\checkmark^{b[n]},\mathcal{G})\\
&\leq&\mbox{E}_{Z_\checkmark^{b[n]},\mathcal{G}}\log|S'(Z_\checkmark^{b[n]},\mathcal{G})|\label{eq:uniform}
\end{eqnarray}
We used functional dependence in (\ref{eq:fd}). Given $Z_\checkmark^{b[n]}$ and $\mathcal{G}$, define $S'(Z_\checkmark^{b[n]},\mathcal{G})$ as the set of feasible codewords, or equivalently the set of feasible $Z_\times^{[n]}$ (because of functional dependence). In (\ref{eq:uniform}) we used the fact that the uniform distribution maximizes entropy. 

For the aligned images arguments, it is more convenient to index the aligned image sets by $Z_\times^{[n]}$ instead of $Z_\checkmark^{b[n]}$ values. This is accomplished as follows.
\begin{eqnarray}
&&H(Z_\times^{[n]}|Z_\checkmark^{b[n]},\mathcal{G})\nonumber\\
&\leq&\mbox{E}_{Z_\checkmark^{b[n]},\mathcal{G}}\log|S'(Z_\checkmark^{b[n]},\mathcal{G})|\\
&=&\mbox{E}_{\mathcal{G}}\sum_{z_\checkmark^{b[n]}\in\mathcal{Z}_\checkmark^{[n]}}\mathbb{P}(Z_\checkmark^{b[n]}=z_\checkmark^{b[n]}|\mathcal{G})\log|S'(z_\checkmark^{b[n]},\mathcal{G})|\\
&=&\mbox{E}_{\mathcal{G}}\sum_{z_\checkmark^{b[n]}\in\mathcal{Z}_\checkmark^{[n]}}~~~~
\sum_{z_\times^{[n]}\in\mathcal{Z}_\times^{[n]}: Z_\checkmark^{b[n]}(z_\times^{[n]},\mathcal{G})=z_\checkmark^{b[n]}}
\mathbb{P}(Z_\times^{[n]}=z_\times^{[n]}|\mathcal{G})\log|S'(z_\checkmark^{b[n]},\mathcal{G})|\\
&=&\mbox{E}_{\mathcal{G}}\sum_{z_\checkmark^{b[n]}\in\mathcal{Z}_\checkmark^{[n]}}~~~~
\sum_{z_\times^{[n]}\in\mathcal{Z}_\times^{[n]}: Z_\checkmark^{b[n]}(z_\times^{[n]},\mathcal{G})=z_\checkmark^{b[n]}}
\mathbb{P}(Z_\times^{[n]}=z_\times^{[n]})\log|S'(z_\checkmark^{b[n]},\mathcal{G})|\label{eq:ind}\\
&=&\mbox{E}_{\mathcal{G}}\sum_{z_\checkmark^{b[n]}\in\mathcal{Z}_\checkmark^{[n]}}~~~~
\sum_{z_\times\in\mathcal{Z}_\times^{[n]}: Z_\checkmark^{b[n]}(z_\times^{[n]},\mathcal{G})=z_\checkmark^{b[n]}}
\mathbb{P}(Z_\times^{[n]}=z_\times^{[n]})\log|S(z_\times^{[n]},\mathcal{G})|\\
&=&\mbox{E}_{\mathcal{G}}\sum_{z_\times^{[n]}\in\mathcal{Z}_\times^{[n]}}\mathbb{P}(Z_\times^{[n]}=z_\times^{[n]})\log|S(z_\times^{[n]},\mathcal{G})|\\
&=&\sum_{z_\times^{[n]}\in\mathcal{Z}_\times^{[n]}}\mathbb{P}(Z_\times^{[n]}=z_\times^{[n]})\mbox{E}_{\mathcal{G}}\log|S(z_\times^{[n]},\mathcal{G})|\\
&\leq&\sum_{z_\times^{[n]}\in\mathcal{Z}_\times^{[n]}}\mathbb{P}(Z_\times^{[n]}=z_\times^{[n]})\log\mbox{E}_{\mathcal{G}}|S(z_\times^{[n]},\mathcal{G})|\label{eq:Jensens}\\
&\leq&\max_{z_\times^{[n]}\in\mathcal{Z}_\times^{[n]}}\log\mbox{E}_{\mathcal{G}}|S(z_\times^{[n]},\mathcal{G})|\\
&=&\log\mbox{E}_{\mathcal{G}}|S(\nu^{[n]},\mathcal{G})|\label{eq:nu}\\
&=&\log\left(\sum_{\lambda\in\mathcal{Z}_\times^{[n]}}\mathbb{P}(\lambda^{[n]}\in S(\nu^{[n]},\mathcal{G}))\right)
\end{eqnarray}
where $\mathcal{Z}_\checkmark^{[n]}$ and $\mathcal{Z}_\times^{[n]}$ are defined as the support of the random variables $Z_\checkmark^{b[n]}$ and $Z_\times^{[n]}$, respectively. In (\ref{eq:ind}) we used the fact that $Z_\times^{[n]}$ is independent of $\mathcal{G}$. This is because it depends only on the codewords, which are chosen independent of the realizations of $\mathcal{G}$. The aligned image set $S(Z_\times^{[n]}, \mathcal{G})$ is defined as follows.
\begin{eqnarray}
S(Z_\times^{[n]},\mathcal{G})&=&\{\lambda^{[n]} \in\mathcal{Z}_\times^{[n]} \mbox{ such that } Z_\checkmark^{b[n]}(\lambda,\mathcal{G})=Z_\checkmark^{b[n]}(Z_\times,\mathcal{G})\}
\end{eqnarray}
Jensen's inequality was used to obtain (\ref{eq:Jensens}). Equation (\ref{eq:nu}) is based on the following definition of $\nu^{[n]}$,
\begin{eqnarray}
\nu^{[n]}=\arg \max_{z_\times^{[n]}\in\mathcal{Z}_\times^{[n]}}\log\mbox{E}_{\mathcal{G}}|S(z_\times^{[n]},\mathcal{G})|.
\end{eqnarray}
\subsubsection{Bounding the Probability of Alignment $\mathbb{P}(\lambda^{[n]}\in S(\nu^{[n]},\mathcal{G}))$}
Consider two distinct realizations of $Z_\times^{[n]}$, denoted by $\lambda^{[n]}$ and $\nu^{[n]}$. We wish to bound the probability that they align, i.e., that they produce the same $Z_\checkmark^{b[n]}$. Let us denote the corresponding codewords realizations $\bar{X}_{i,j}^{[n]}$  by $\lambda_{i,j}^{[n]}$ and $\nu_{i,j}^{[n]}$, respectively. 
\begin{eqnarray}
\lambda^{[n]} &=&(L_{1\times}^{[n]}(\lambda_{1,l'_1},\lambda_{2,0}),L_{2\times}^{[n]}(\lambda_{2,l'_2},\lambda_{3,0})\cdots,L_{m\times}^{[n]}(\lambda_{m,l'_{m}},\lambda_{1,0}))\\
&\triangleq&(\lambda_1^{[n]},\lambda_2^{[n]},\cdots,\lambda_{m}^{[n]})\\
\nu^{[n]} &=&(L_{1\times}^{[n]}(\nu_{1,l'_1},\nu_{2,0}),L_{2\times}^{[n]}(\nu_{2,l'_2},\nu_{3,0})\cdots,L_{m\times}^{[n]}(\nu_{m,l'_m},\nu_{1,0}))\\
&\triangleq&(\nu_1^{[n]},\nu_2^{[n]},\cdots,\nu_{m}^{[n]})
\end{eqnarray}
As required for the aligned images argument, our goal in this section is to bound $\mathbb{P}(\lambda\in S(\nu^{[n]},\mathcal{G}))$ from above, with an expression involving the $|\lambda_i(t)-\nu_i(t)|$ terms.

\noindent Given $\mathcal{G}$, if $\lambda^{[n]}\in S(\nu^{[n]},\mathcal{G})$, then 
\begin{eqnarray}
Z^{b[n]}_{\checkmark}(\lambda^{[n]},\mathcal{G})&=&Z^{b[n]}_{\checkmark}(\nu^{[n]},\mathcal{G})
\end{eqnarray}
i.e.,
\begin{eqnarray}
\lefteqn{(L^{b[n]}_{1{\checkmark}}(\lambda_{1,0},\lambda_{1,l_1}),L^{b[n]}_{2{\checkmark}}(\lambda_{2,0},\lambda_{2,l_2}),\cdots,L^{b[n]}_{m{\checkmark}}(\lambda_{m,0},\lambda_{m,l_{m}}))}\\
&=&(L^{b[n]}_{1{\checkmark}}(\nu_{1,0},\nu_{1,l_1}),L^{b[n]}_{2{\checkmark}}(\nu_{2,0},\nu_{2,l_2}),\cdots,L^{b[n]}_{m{\checkmark}}(\nu_{m,0},\nu_{m,l_{m}})).
\end{eqnarray}
So for all $t\in[n]$, and for all $i\in[m]$, we have,
\begin{eqnarray}
\lfloor g_{i,0}(t)\lambda_{i,0}(t)\rfloor+\lfloor g_{i,l_i}(t)\lambda_{i,l_i}(t)\rfloor&=&\lfloor g_{i,0}(t)\nu_{i,0}(t)\rfloor+\lfloor g_{i,l_i}(t)\nu_{i,l_i}(t)\rfloor\\
\implies \lfloor g_{i,0}(t)\lambda_{i,0}(t)\rfloor-\lfloor g_{i,0}(t)\nu_{i,0}(t)\rfloor
&=&\underbrace{\lfloor g_{i,l_i}(t)\nu_{i,l_i}(t)\rfloor-\lfloor g_{i,l_i}(t)\lambda_{i,l_i}(t)\rfloor}_{\triangleq a_i(t)}
\end{eqnarray}
\begin{eqnarray}
 g_{i,0}(t)\left(\lambda_{i,0}(t)-\nu_{i,0}(t)\right)&\in&(a_i(t)-2,a_i(t)+2)
\end{eqnarray}
Thus, conditioned on any given value of $g_{i,l_i}(t)$, alignment of $\lambda^{[n]}$ and $\nu^{[n]}$ requires that $g_{i,0}(t)$ must take values in an interval of length less than or equal to $4/|\lambda_{i,0}(t)-\nu_{i,0}(t)|$.\footnote{If $\lambda_{i,0}(t)=\nu_{i,0}(t)$ then the interval is of infinite length, which renders the constraint inactive.} Similarly, conditioned on any given value of $g_{i,0}(t)$, alignment requires that $g_{i,l_i}(t)$ must take values in an interval of length less than or equal to $4/|\lambda_{i,l_i}(t)-\nu_{i,l_i}(t)|$. From each pair of channels $g_{i,0}(t)$ and $g_{i,l_i}(t)$, let us define $\bar{g}_i(t)$ as the one that corresponds to the smaller interval, while the other is identified as $\bar{g}_i^c(t)$. Let us also define $B_{i,j}(t)$ which will be useful at a later stage of this proof.
Define
\begin{eqnarray}
B_{i,j}(t)&\triangleq&\left\{
\begin{array}{ll}
\max\Big(|\lambda_{i,l_i}(t)-\nu_{i,l_i}(t)|, |\lambda_{i,0}(t)-\nu_{i,0}(t)|\Big)&\mbox{ if } i=j\\
\max\Big(|\lambda_{i,l'_i}(t)-\nu_{i,l'_i}(t)|, |\lambda_{j,0}(t)-\nu_{j,0}(t)|\Big)&\mbox{ if } i\neq j
\end{array}
\right.\\
(\bar{g}_i(t),\bar{g}_i^c(t))&\triangleq&\left\{
\begin{array}{ll}
(g_{i,0}(t),g_{i,l_i}(t))&\mbox{ if } B_{i,i}(t)= |\lambda_{i,0}(t)-\nu_{i,0}(t)|\\
(g_{i,l_i}(t),g_{i,0}(t))&\mbox{ if }  B_{i,i}(t)\neq|\lambda_{i,0}(t)-\nu_{i,0}(t)|
\end{array}
\right.
\end{eqnarray}
Thus, $\forall i\in[m], \forall t\in[n]$, for $\lambda^{[n]}\in S(\nu^{[n]},\mathcal{G})$, it must be true that conditioned on any value of $\bar{g}_i^c(t)$,  the bounded density random variable $\bar{g}_i(t)$  takes values in an interval $\delta_i(t)$ of length $4/B_{i,i}(t)$. Therefore, the bounded density assumption on $\mathcal{G}$, leads to the following bound on the probability of alignment.
\begin{eqnarray}
\mathbb{P}(\lambda^{[n]}\in S(\nu^{[n]},\mathcal{G}))&\leq&\idotsint_{*}f(\bar{g}^c_*)\left(\idotsint_{\bar{g}_*\in \delta_*}f(\bar{g}_*\mid \bar{g}^c_*)d\bar{g}_*\right)d\bar{g}^c_*\\
&\leq&\idotsint_{*}f(\bar{g}^c_*)\left(\prod_{i\in[m]}\prod_{\substack{t\in[n]\\ B_{i,i}(t)\neq 0} }\frac{4f_{\max}}{B_{i,i}(t)}\right)d\bar{g}^c_*\\
&=&\prod_{i\in[m]}\prod_{\substack{t\in[n]\\ B_{i,i}(t)\neq 0} }\frac{4f_{\max}}{B_{i,i}(t)}\\
&\leq&(4f_{\max})^{mn}\prod_{i\in[m]}\prod_{\substack{t\in[n]} }\frac{1}{B_{i,i}^+(t)}\label{eq:Bchange}
\end{eqnarray}
where $B_{i,j}^+(t)\triangleq \max(1, B_{i,j}(t))$, i.e., when $B_{i,j}(t)=0$ then $B_{i,j}^+(t)=1$. (\ref{eq:Bchange}) holds because $f_{\max}\geq 1$. Thus, we have a bound on $\mathbb{P}(\lambda^{[n]}\in S(\nu^{[n]},\mathcal{G}))$ in terms of $|\lambda_{i,j}(t)-\nu_{i,j}(t)|$ terms. Recall that $\lambda_{i,j}(t)$ and $\nu_{i,j}(t)$ are the  realizations of codeword symbols $\bar{X}_{i,j}(t)$. However, for the aligned images argument, we need the bound in terms of $|\lambda_i(t)-\nu_i(t)|$ terms, where $\lambda_i(t)$ and $\nu_i(t)$ are the corresponding realizations of the elements of $Z_\times$. This is accomplished through a novel argument as follows.

For all $i\in[m]$, and $\forall t\in[n]$,
\begin{eqnarray}
\lambda_i(t)-\nu_i(t)=\lfloor h_{i,l'_i}(t)\lambda_{i,l'_i}(t)\rfloor+\lfloor h_{i+1,0}(t)\lambda_{i+1,0}(t)\rfloor -\lfloor h_{i,l'_i}(t)\nu_{i,l'_i}(t)\rfloor-\lfloor h_{i+1,0}(t)\nu_{i+1,0}(t)\rfloor
\end{eqnarray}
\begin{eqnarray}
\implies |\lambda_i(t)-\nu_i(t)|& \leq&2\Delta_2\max\Big(|\lambda_{i,l'_i}(t)-\nu_{i,l'_i}(t)|, |\lambda_{i+1,0}(t)-\nu_{i+1,0}(t)|\Big)+2\\
&=&2\Delta_2B_{i,i+1}(t)+2\label{eq:Bbound}
\end{eqnarray}

In order to go from $B_{i,i}^+(t)$ terms in (\ref{eq:Bchange}) to $|\lambda_i(t)-\nu_i(t)|$ terms, we wish to  replace the $B_{i,i}^+(t)$ terms with $B_{i,i+1}^+(t)$ terms. To this end, define
\begin{eqnarray}
i^*(t)&=&\arg \max_i B^+_{i,i}(t) \label{eq:critical}
\end{eqnarray}
which then implies
\begin{eqnarray}
B_{i^*,i^*+1}^+(t)&\leq&B^+_{i^*,i^*}(t)\\
B_{i^*+2,i^*+3}^+(t)&\leq&B^+_{i^*+2,i^*+2}(t)B^+_{i^*+3,i^*+3}(t)\\
B_{i^*+4,i^*+5}^+(t)&\leq&B^+_{i^*+4,i^*+4}(t)B^+_{i^*+5,i^*+5}(t)\\
&\vdots&\\
B_{i^*+m-1,i^*+m}^+(t)&\leq&B^+_{i^*+m-1,i^*+m-1}(t)B^+_{i^*+m,i^*+m}(t)
\end{eqnarray}
The remaining $B_{i,i+1}^+(t)$ terms are bounded as follows.
\begin{eqnarray}
B^+_{i^*+1,i^*+2}(t)&\leq&\bar{P}\\
B^+_{i^*+3,i^*+4}(t)&\leq&\bar{P}\\
&\vdots&\\
B^+_{i^*+m-2,i^*+m-1}(t)&\leq&\bar{P}
\end{eqnarray}
Substituting into (\ref{eq:Bchange}) we have,
\begin{eqnarray}
\mathbb{P}(\lambda^{[n]}\in S(\nu^{[n]},\mathcal{G}))&\leq&\bar{P}^{n(m-1)/2}(4f_{\max})^{mn}\prod_{i\in[m]}\prod_{t\in[n]}\frac{1}{B_{i,i+1}^+(t)}
\end{eqnarray}
and further substituting from (\ref{eq:Bbound}) we have
\begin{eqnarray}
\mathbb{P}(\lambda^{[n]}\in S(\nu^{[n]},\mathcal{G}))&\leq&\bar{P}^{n(m-1)/2}(4f_{\max})^{mn}\nonumber\\
&&\prod_{i\in[m]}\left(\prod_{\substack{t\in[n]\\ |\lambda_i(t)-\nu_i(t)|>2}}\frac{2\Delta_2}{|\lambda_i(t)-\nu_i(t)|-2}\right)\left(\prod_{\substack{t\in[n]\\ |\lambda_i(t)-\nu_i(t)|\leq 2}}1\right)\nonumber\\
&\leq&\bar{P}^{n(m-1)/2}(8\Delta_2f_{\max})^{mn}\prod_{i\in[m]}\prod_{\substack{t\in[n]\\ |\lambda_i(t)-\nu_i(t)|>2}}\frac{1}{|\lambda_i(t)-\nu_i(t)|-2}\label{eq:Delta2}
\end{eqnarray}
(\ref{eq:Delta2}) holds because $\Delta_2\geq 1$. Thus, we have our desired bound.
\subsubsection{Bounding the average size of the aligned image set, $\mbox{E}_\mathcal{G}|S(\nu^{[n]},\mathcal{G})|$}
\begin{align}
\mbox{E}_\mathcal{G}|S(\nu^{[n]},\mathcal{G})|&=\sum_{\lambda^{[n]}\in\mathcal{Z}_\times^{[n]}}\mathbb{P}(\lambda^{[n]}\in S(\nu^{[n]},\mathcal{G}))
\end{align}
\begin{align}
&\leq\sum_{\lambda^{[n]}\in\mathcal{Z}_\times^{[n]}}\bar{P}^{n(m-1)/2}(8\Delta_2f_{\max})^{mn}\prod_{i\in[m]}\left(\prod_{\substack{t\in[n]\\ |\lambda_i(t)-\nu_i(t)|>2}}\frac{1}{|\lambda_i(t)-\nu_i(t)|-2}\times \prod_{\substack{t\in[n]\\ |\lambda_i(t)-\nu_i(t)|\leq 2}}1\right)\label{eq:sumprod}\\
&\leq\bar{P}^{n(m-1)/2}(8\Delta_2f_{\max})^{mn}\prod_{i\in[m]}\prod_{t\in[n]}\left(\sum_{\substack{\lambda_i(t)\in[\hat{P}]\\ |\lambda_i(t)-\nu_i(t)|>2}}\frac{1}{|\lambda_i(t)-\nu_i(t)|-2}+ \sum_{\substack{\lambda_i(t)\in[\hat{P}]\\ |\lambda_i(t)-\nu_i(t)|\leq 2}}1\right)\label{eq:prodsum}\\
&\leq \bar{P}^{n(m-1)/2}(8\Delta_2f_{\max})^{mn}\prod_{i\in[m]}\prod_{t\in[n]}\left(2\sum_{p\in[\hat{P}]}\frac{1}{p}+ 5\right)\\
&\leq \bar{P}^{n(m-1)/2}(8\Delta_2f_{\max})^{mn}\prod_{i\in[m]}\prod_{t\in[n]}\left(2+2\log(\hat{P})+ 5\right)\label{sigm}\\
&= \bar{P}^{n(m-1)/2}(8\Delta_2f_{\max})^{mn}(7+2\log(\hat{P}))^{mn}
\end{align}
where $\hat{P}=3+\lfloor2\Delta_2\bar{P}\rfloor$. (\ref{eq:prodsum}) follows from interchange of the summation and the product.\footnote{ Note that for the arbitrary functions $f_1(x),f_2(x),\cdots,f_n(x)$ and the arbitrary sets of numbers $S_1,S_2,\cdots,S_n$ we have,
\begin{eqnarray}
\sum_{a_1\in S_1,a_2\in S_2,\cdots,a_n\in S_n}\prod_{t=1}^nf_t(a_t)&=&\prod_{t=1}^n\sum_{a_t\in S_t}f_t(a_t)
\end{eqnarray}} 
(\ref{sigm}) is true as the partial sum of harmonic series can be bounded above by logarithmic
function, i.e., $\sum_{i=1}^n\frac{1}{i}\le1+\log{n}$.

\subsubsection{Contradiction}
Substituting into (\ref{eq:nu}), we have
\begin{eqnarray}
H(Z_\times^{[n]})-H(Z_\checkmark^{b[n]}|\mathcal{G})&\leq& \log\mbox{E}_\mathcal{G}|S(\nu^{[n]},\mathcal{G})|\\
&\leq&\log\left(\bar{P}^{n(m-1)/2}(8\Delta_2f_{\max})^{mn}(7+2\log(\hat{P}))^{mn}\right)\\
&=&\frac{(m-1)}{2}n\log(\bar{P})+no(\log(\bar{P}))
\end{eqnarray}

\noindent Comparing with (\ref{eq:general}) we have a general bound on the symmetric DoF per user, $\alpha$,
\begin{eqnarray}
\Big(\alpha m+(2\alpha-1)l_\Sigma\Big)n\log(\bar{P})&\leq&\frac{(m-1)}{2}n\log(\bar{P})
\end{eqnarray}
\begin{eqnarray}
\implies \alpha&\leq&\left(\frac{1}{2}\right)\left(1-\frac{1}{m+2l_\Sigma}\right)\label{eq:mainbound}
\end{eqnarray}

 \section{Conclusion}
A DoF bound sensitive to network coherence time was obtained. This was accomplished by a novel adaptation ((\ref{eq:critical})-(\ref{eq:Delta2})) of the aligned image sets bound, and closes several open problems noted previously by  Naderi and Avestimehr in \cite{Naderi_Avestimehr} and by Gou et al. in \cite{Gou_TIM}.

\appendix\section{Achieving $4/9$ DoF per User in the Network of Figure \ref{fig:graphs}(a)}\label{4/9ach}
Consider three channel uses. For any $i\in[7]$ user $i$'s message $W_i$ is split into messages $W_{ic}$ and $W_{ip}$, representing common message and private message, respectively. The common message $W_{ic}$ is encoded into the symbol $X_{ic}$ and may be decoded by several receivers while the private message $W_{ip}$ is encoded to $X_{ip}$ and is intended to be decoded by the $i$-th receiver.  The codeword $X_{ip}$ carries $1$ DoF for any $i\in[7]$ while the codeword $X_{ic}$ carries $\frac{1}{3}$ DoF. $X_{ip}$ and $X_{ic}$  are transmitted with powers
\begin {eqnarray}
\mbox{E}{|X_{ip}|}^2&=&0.5\\
\mbox{E}{|X_{ic}|}^2&=&0.5
\end{eqnarray}
Since the reduced graph $\mathbb{G}_r$  has an odd cycle with the length $m=3$, it is $3$-colorable. Instead of the three colors, consider the three vectors ${\bf e}_1=(1,0,0)^T$, ${\bf e}_2=(0,1,0)^T$ and ${\bf e}_3=(0,0,1)^T$. We assign either ${\bf e}_1$, ${\bf e}_2$ or ${\bf e}_3$ to each vertex of $\mathbb{G}_r$, such that no two conflicting vertices are assigned the same ${\bf e}_i$, e.g., assign ${\bf e}_1$ to vertex $V_1$ of $\mathbb{G}_r$ which corresponds to the alignment set $\mathcal{A}_1$ in the alignment graph $\mathbb{G}_a$, i.e., messages $W_{1p}$ and $W_{2p}$. Moreover, assign ${\bf e}_2$ to messages $W_{4p}$ and $W_{7p}$ and ${\bf e}_3$ to messages $W_{3p}$, $W_{5p}$ and $W_{6p}$. In the first channel use, all messages that are assigned the vector ${\bf e}_1$ are transmitted. Similarly, in the second channel use, all messages that are assigned the vector ${\bf e}_2$ are transmitted and in the third channel use, all messages that are assigned the vector ${\bf e}_3$ are transmitted. 

Now, we can make the following observation. Let us denote the $i$-th transmitter and $i$-th receiver by $T_i$ and $R_i$, respectively. For any $i\in[7]$, $R_i$ receives signals from the transmitters in  two of the three time slots while it does not receive any signal in the remaining one time slot. For instance, $R_1$ sees the message from $T_1$ at the first time slot and the messages from the $T_3$ and $T_5$ at the third time slot while no messages from the transmitters are received at the second time slot. In the other words, the received signals at the receivers only span two dimensions out of three possible dimensions. So, we assign the vector ${\bf e}_4=(1,1,1)^T$ to all the remaining messages, i.e., $W_{1c},W_{2c},\cdots,W_{7c}$. Each codeword $X_{ic}$ corresponding to the message $W_{ic}$ is transmitted in all the three time slots. In the other words, the transmitted signals are,
\begin{eqnarray}
{\bf X}_{i}&=&\left\{\begin{matrix}
{\bf e}_1X_{ip}+{\bf e}_4X_{ic} & i\in\{1,2\} \\ 
{\bf e}_2X_{ip}+{\bf e}_4X_{ic} & i\in\{4,7\} \\ 
{\bf e}_3X_{ip}+{\bf e}_4X_{ic} & i\in\{3,5,6\} \\ 
\end{matrix}\right.
\end{eqnarray}
where the received signals are shown in (\ref{chm}). Now we claim that each receiver $R_i$ can decode its own messages $X_{ic}$ and $X_{ip}$. Consider the received signal at the first receiver,
\begin{eqnarray}
Y_1(1)&=&\sqrt{P}\Big(G_{11}(1)X_{1p}+G_{11}(1)X_{1c}+G_{13}(1)X_{3c}+G_{15}(1)X_{5c}\Big)+Z_1(1)\\
Y_1(2)&=&\sqrt{P}\Big(G_{11}(2)X_{1c}+G_{13}(2)X_{3c}+G_{15}(2)X_{5c}\Big)+Z_1(2)\\
Y_1(3)&=&\sqrt{P}\Big(G_{13}(3)X_{3p}+G_{15}(3)X_{5p}+G_{11}(3)X_{1c}+G_{13}(3)X_{3c}+G_{15}(3)X_{5c}\Big)+Z_1(3)
\end{eqnarray}
Now we claim that $R_1$ can decode the messages $W_{1c},W_{3c},W_{5c}$ as a MAC in the second channel use.\footnote{Note that each of the messages $W_{1c},W_{3c},W_{5c}$ has $\frac{1}{3}$ DoF and the received signal in the second channel use is
\begin{eqnarray}
Y_1(2)&=&\sqrt{P}\left(G_{11}(2)X_{1c}+G_{13}(2)X_{3c}+G_{15}(2)X_{5c}\right)+Z_1(2).
\end{eqnarray}
As $\mbox{E}{|X_{ic}|}^2=0.5$ for any $i\in\{1,3,5\}$, $R_1$ can decode the messages $W_{1c},W_{3c},W_{5c}$.}
 Moreover, after decoding  the messages $W_{1c},W_{3c},W_{5c}$, $R_1$ can reconstruct the codewords $X_{1c},X_{3c},X_{5c}$, subtract them from the received signal in the first channel use and decode the desired message $W_{1p}$. Therefore, $R_1$ can decode its own desired messages. Similarly all the receivers can decode their own desired messages resulting in total $\frac{4}{3}$ DoF in three channel uses. Note that $\frac{4}{9}$ is achievable in the interference networks in Figures \ref{Fig2} and \ref{Fig3} similarly.

\bibliographystyle{IEEEtran}
\bibliography{Thesis}

\end{document}